\documentclass[prl,twocolumn,showpacs,superscriptaddress,floatfix,longbibliography]{revtex4-1}
\usepackage{graphicx,amsfonts,amssymb,amsmath,xspace,dsfont}
\usepackage[colorlinks=true,citecolor=green,linkcolor=red,urlcolor=blue]{hyperref}
\usepackage{subfigure}
\usepackage{calligra}

\usepackage{color}
\definecolor{g}{rgb}{.1,0.4,.1} % {.0,0.7,.5}
\definecolor{b}{rgb}{0,0.2,1}
\definecolor{rouge}{rgb}{0.82,0.,0.}
\definecolor{vert}{rgb}{0.,0.82,0.}
\definecolor{orange}{rgb}{1,0.5,0.}
\definecolor{bleu}{rgb}{0.,0.,0.82}
\definecolor{m}{rgb}{0.82,0.,0.82}
\definecolor{vert2}{rgb}{0.,0.5,0.}
\definecolor{rougeclair}{rgb}{1.0,0.7,0.7}

\newcommand{\beq}{\begin{equation}}
\newcommand{\be}{\begin{equation}}
\newcommand{\beqn}{\begin{eqnarray}}
\newcommand{\eeq}{\end{equation}}
\newcommand{\ee}{\end{equation}}
\newcommand{\eeqn}{\end{eqnarray}}

\newcommand{\bem}{\begin{pmatrix}}

\newcommand{\eem}{\end{pmatrix}}

\newlength{\ldag}
\newcommand{\bra}[1]{\langle#1|}
\newcommand{\ket}[1]{|#1\rangle}

\newcommand{\mN}{\mathcal{N}}

\newcommand{\id}{\mathds{1}}

\newcommand{\Pp}{B_p}

\newcommand{\Pe}{L_l}
\newcommand{\C}{\mathcal{R}}

\begin{document}

\title{Wegner-Wilson loops in string nets}

\author{Anna Ritz-Zwilling}
\email{anna.ritz@u-psud.fr}
\affiliation{Sorbonne Universit\'e, CNRS, Laboratoire de Physique Th\'eorique de la Mati\`ere Condens\'ee, LPTMC, F-75005 Paris, France}

\author{Jean-No\"el Fuchs}
\email{fuchs@lptmc.jussieu.fr}
\affiliation{Sorbonne Universit\'e, CNRS, Laboratoire de Physique Th\'eorique de la Mati\`ere Condens\'ee, LPTMC, F-75005 Paris, France}

\author{Julien Vidal}
\email{vidal@lptmc.jussieu.fr}
\affiliation{Sorbonne Universit\'e, CNRS, Laboratoire de Physique Th\'eorique de la Mati\`ere Condens\'ee, LPTMC, F-75005 Paris, France}

%%%%%%%%%%%%%%%%%%%%%%%%%%%%%%%%%
%%%%%%%%%%%%%%%%%%%%%%%%%%%%%%%%%
\begin{abstract}
We study the Wegner-Wilson loops in the string-net model of Levin and Wen in the presence of a string tension. The latter is responsible for a phase transition from a topological deconfined phase (weak tension) to a trivial confined phase (strong tension). We analyze the behavior of all Wegner-Wilson loops in both limiting cases for an arbitrary input theory of the string-net model. Using a fluxon picture, we compute perturbatively the first contributions to a perimeter law in the topological phase as a function of the quantum dimensions. In the trivial phase, we find that Wegner-Wilson loops obey a modified area law, in agreement with a recent mean-field approach. 
\end{abstract}

\maketitle
%%%%%%%%%%%%%%%%%%%%%%%%%%%%%%%%%
%%%%%%%%%%%%%%%%%%%%%%%%%%%%%%%%%
%
%%%%%%%%%%%%%%%%%%%
%%%%%%%%%%%%%%%%%%%
\noindent\emph{Introduction}---
%%%%%%%%%%%%%%%%%%%
%%%%%%%%%%%%%%%%%%%
%
Lattice gauge theories were introduced by Wegner \cite{Wegner71}  in the early 1970s to study classical phase transitions that cannot be described by a local order parameter. Shortly after, Wilson proposed a lattice version of quantum chromodynamics to describe the quark confinement \cite{Wilson74}, hence extending Wegner's work based on the $\mathbb{Z}_2$ gauge group to arbitrary gauge groups (see also Refs.~\cite{Kogut75,Kogut79}). In the absence of matter (pure gauge theories), one generally distinguishes between two phases characterized by the behavior of nonlocal gauge-invariant correlation functions defined along a closed contour, dubbed Wegner-Wilson loops. In the confined (strong-interaction) phase, the expectation value of these loops in the ground state decays as ${\rm e}^{-\# A}$ (area law), whereas in the deconfined (weak-interaction) phase they behave as ${\rm e}^{-\# L}$ (perimeter law), where $A$ and $L$ denote the area and the perimeter of the loop, respectively. When matter is included, the Wegner-Wilson loop features a perimeter law in both phases and another diagnostic of the transition is required \cite{Gregor11}.

In two dimensions, lattice gauge theories are of special interest since they may host exotic excitations known as anyons \cite{Leinaas77,Wilczek82}. The latter have drawn much attention tin recent decades because of their potential use for topological quantum computation~\cite{Kitaev03,Ogburn99,Freedman03,Preskill_HP,Nayak08,Wang_book}, and they are considered as a hallmark of systems with topological order. During the last three decades, the concept of topological order has become central in condensed matter physics, and several models have been proposed to generate topological phases of matter (see Ref.~\cite{Wen17} for a recent review). Among them, the string-net model introduced by Levin and Wen~\cite{Levin05} is particularly interesting since it goes beyond lattice gauge theories and allows one to build a large class of  topological phases. This model  is closely related to the Turaev-Viro model~\cite{Turaev92,Kadar10,Kirillov11,Koenig10} and can be seen as a discrete version of some topological quantum field theories~\cite{Witten89,Moore89}. 

In this article, we investigate the behavior of Wegner-Wilson loops in the string-net model \cite{Levin05} in the presence of a string tension. This tension is responsible for a phase transition between a deconfined topological phase (weak tension) and a confined trivial phase (strong tension). In the deconfined phase, we compute perturbatively the expectation values of the Wegner-Wilson loops in the ground state and we show that they all obey a perimeter law. In the confined phase, using perturbative and mean-field approaches, we obtain either  a usual or a modified area law depending on the loop considered. We also prove that Wegner-Wilson loops associated with Abelian fluxons commute with the Hamiltonian and remain constant for any strength of the string tension, indicating a complete deconfinement of these excitations.\\

%%%%%%%%%%%%%%%%%%%
%%%%%%%%%%%%%%%%%%%
\noindent\emph{The Levin-Wen Model}---
%%%%%%%%%%%%%%%%%%%
%%%%%%%%%%%%%%%%%%%
%
%
%%%%%%%%%%%%%
\begin{figure}[t]
\includegraphics[width=0.95\columnwidth]{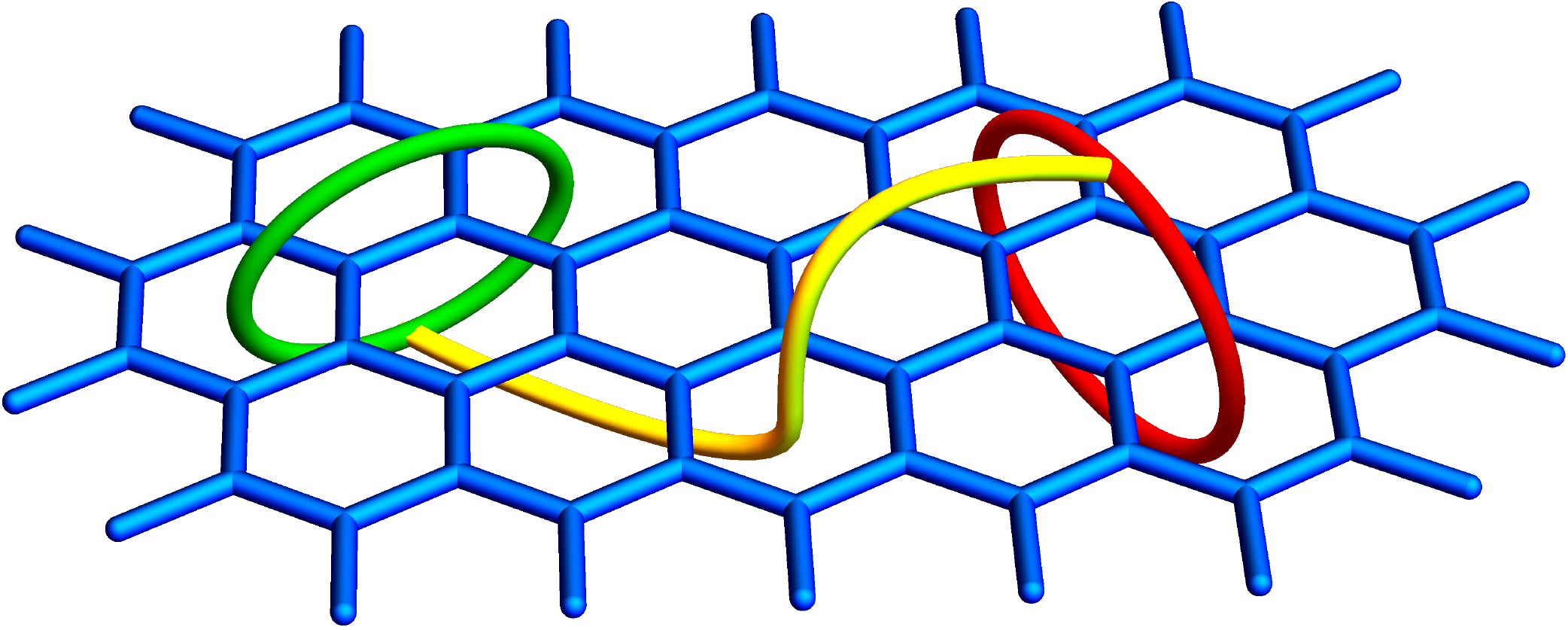}
\caption{Pictorial representation of a five-fluxon state for a theory with three nontrivial strings $s=r,g,y$ (red, green, and yellow). The trivial string $s=1$ is not represented.}
\label{fig:example}
\end{figure}
%%%%%%%%%%%%%
%!TEX encoding = UTF-8 Unicode
%
%
In the string-net model introduced by Levin and Wen \cite{Levin05}, microscopic degrees of freedom are strings defined on the links of a trivalent graph and obeying a set of rules given by an input theory. Here, we focus on input theories that are unitary modular tensor categories (UMTCs) (see Refs.~\cite{Rowell09,Bonderson_thesis,Wang_book} for an introduction), and we consider the honeycomb lattice as a prototypical trivalent graph. A UMTC is defined by a set of $n_s$ strings obeying fusion and braiding rules \cite{Note}. The trivial string $s=1$ corresponds to the vacuum. Simplest examples of UMTC are the semion  and the Fibonacci theories for which $n_s=2$. 
The Hilbert space $\mathcal{H}$ is spanned by all link (string) configurations satisfying the branching rules that directly stem from the fusion rules. More precisely, a trivalent vertex configuration $(a,b,c)$ is allowed iff the string $c$ belongs to the fusion product of strings $a$ and $b$, i.e., $a\times b$. For any input UMTC, the dimension of $\mathcal{H}$ depends only on the number of vertices. Violations of these branching rules correspond to vertex (charge) excitations that we do not consider here.   The Levin-Wen Hamiltonian is defined by a sum of mutually commuting projectors $\Pp$ defined on each plaquette $p$ (see below). The matrix elements of $\Pp$ in the link basis depend on the input UMTC~\cite{Levin05}. 

Such a construction leads to a topological phase, the excitations of which are identified by determining all closed string operators that commute with the Hamiltonian, known as the Wegner-Wilson loops \cite{Wegner71,Wilson74,Kogut79}. As explained in Ref.~\cite{Levin05}, if $\mathcal C$ has $n_s$ strings, there are $n_s^2$ such operators, each of them corresponding to one type of elementary excitation. The resulting doubled achiral topological phase ${\rm D} \mathcal C=( \mathcal C, \overline{\mathcal C})$ consists of two copies of the input UMTC $\mathcal{C}$ with opposite chiralities, and excitations can be labeled by $(s,s')$, where $s$ and $s'$ are elements of  $\mathcal{C}$ and $\overline{\mathcal C}$, respectively. However, in the absence of branching rules violations, only $n_s$ elementary excitations corresponding to $s'=s$ are present in the system. These achiral excitations $(s,s)$ have a simple interpretation in terms of plaquette excitations (fluxons) and can be represented as a string of type $s$ piercing elementary plaquettes (see Fig.~\ref{fig:example} for illustration). For a given theory, this description allows for a simple counting of the energy-spectrum degeneracies \cite{Simon12,Schulz13,Schulz14,Hu18} and, hence, of the Hilbert space dimension.

\begin{figure}[t]
\includegraphics[width=0.95\columnwidth]{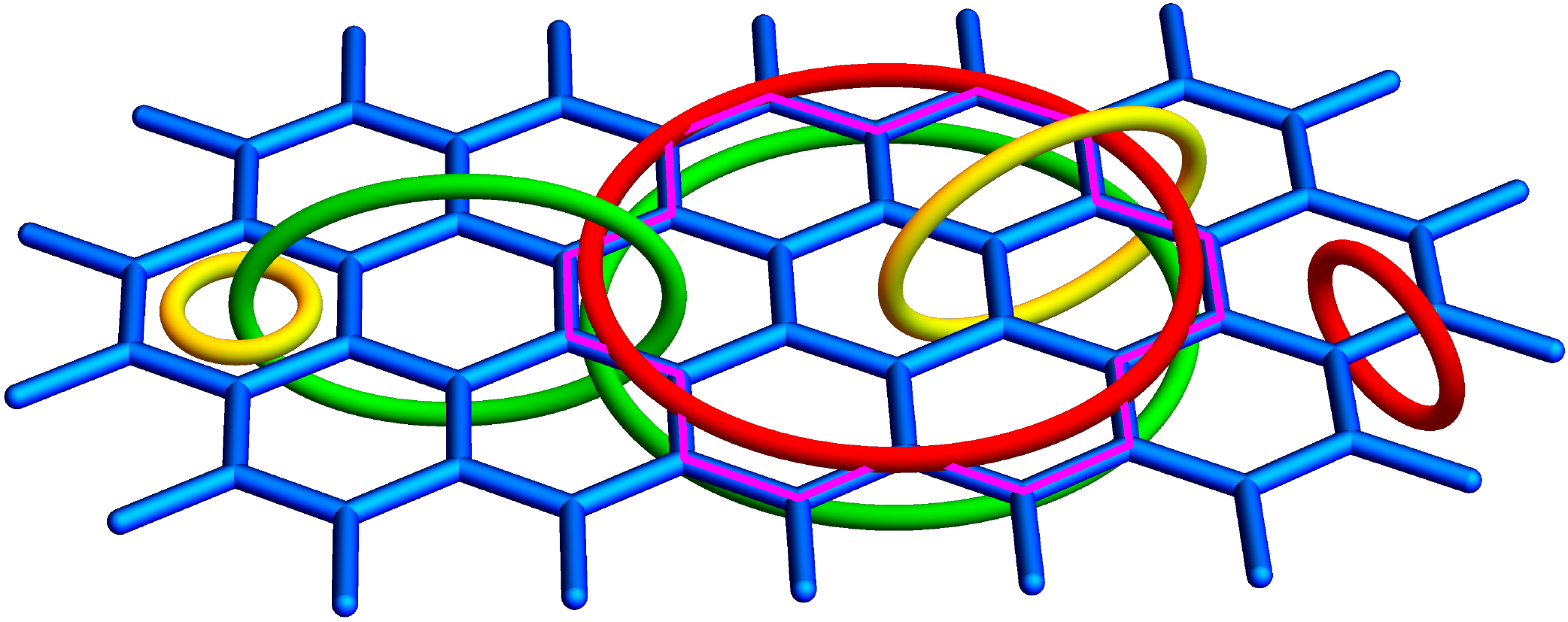}
\caption{Pictorial representation of various operators acting on a four-fluxon state (two $g$ and two $y$). The small yellow loop on the left represents 
$B_p^y$,  the small red loop on the right represents $L_l^r$, and large red and green loops above and below the lattice represent the Wegner-Wilson loop $W^{(r,g)}_\C$. 
The region $\mathcal{R}$ is surrounded  by the magenta line ($L=18, A=7$) in this example. All loops can be smoothly deformed as long as no crossings with either nontrivial strings or links of the lattice are encountered.}
\label{fig:loops}
\end{figure}

\begin{figure}[t]
\includegraphics[width=1\columnwidth]{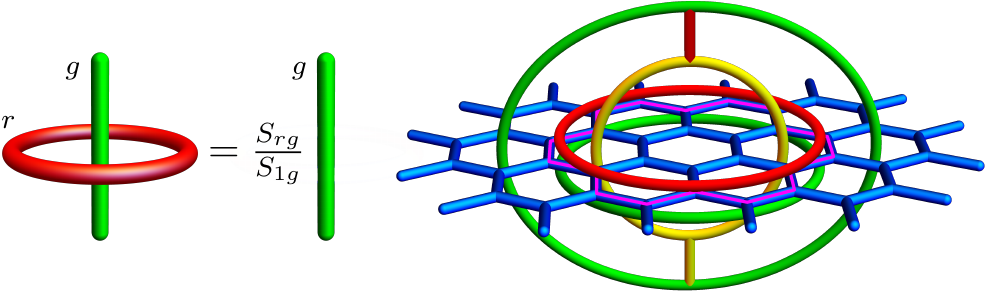}

\caption{Left: a string $g$ is an eigenstate of the operator that injects a closed string $r$ around it with eigenvalue $S_{r g}/S_{1g}$. For a UMTC with $n_s$ strings, $S$ is a symmetric unitary $n_s\times n_s$ matrix. Right: a pair of non-Abelian fluxons  $y$ inside the region $\C$ fuse in $r$ above the lattice and in $y$ below the lattice. For such a state one has \mbox{$\big \langle W^{(r,g)}_\C \big \rangle=\frac{S_{r r}}{S_{1 r}}\frac{S_{g y}}{S_{1 y}}$}. 
}
\label{fig:fundamental}
\end{figure}

The goal of the present work is to study the behavior of the Wegner-Wilson loops in the Levin-Wen model in the presence of a perturbation that plays the role of a string tension and provides dynamics to the fluxons. More precisely, we consider the following Hamiltonian
%
%
%%%%%%%%%%%%%%%%
\begin{equation}
H= - J_\mathrm{p}\sum_p \Pp -J_\mathrm{l}\sum_l \Pe, 
\label{eq:ham}
\end{equation}
%%%%%%%%%%%%%%%%
%
%
where $(J_\mathrm{p},J_\mathrm{l})$ are nonnegative couplings. The original Levin-Wen Hamiltonian is obtained by setting $J_\mathrm{l}=0$. The operator $\Pp$ is the projector onto the state $s=1$  in the pla\-quette~$p$ (fluxon vacuum), and the operator $\Pe$ is the projector onto the state \mbox{$s=1$} in the link~$l$ (string vacuum).  Thus, in the link basis,  $\Pe$ is diagonal and  $\Pp$ is nondiagonal, whereas, in the fluxon basis,  $\Pp$ is diagonal and $\Pe$ is nondiagonal. 
These operators are given by
\be 
\Pp=\frac{1}{D^2} \sum_{s=1}^{n_s} d_s B_p^s, \quad \Pe=\frac{1}{D^2} \sum_{s=1}^{n_s} d_s L_l^s,
\ee
where $d_s$ is the quantum dimension of the string $s$ and \mbox{$D=\sqrt{\sum_s d_s^2}$} is the total quantum dimension of the theory considered. The operator $B_p^s$ injects a closed string $s$ around the \mbox{plaquette $p$} and hence ``measures" the fluxon state in this plaquette, whereas $ L_l^s$ injects a closed string $s$ around the link $l$ and ``measures"  the string state in this link (see Fig.~\ref{fig:loops} for illustration). In this context, measurement refers to the fundamental relation depicted in Fig.~\ref{fig:fundamental} (left), which is reminiscent of the Aharonov-Bohm effect \cite{Aharonov59,Preskill_HP}. All operators $B_p^s$ and $L_l^{s'}$ mutually commute, except when the link $l$ belongs to the plaquette $p$. 

%Using the fundamental relation depicted in Fig.~\ref{fig:fundamental} which is reminiscent of the Aharonov-Bohm effect \cite{Aharonov59}, 

The Hamiltonian (\ref{eq:ham}) has been first introduced by Gils {\it et al.} in the ladder geometry~\cite{Gils09_1,Gils09_3} (see also Refs.~\cite{Ardonne11,Schulz15,Schulz16_2} for related studies). In the honeycomb lattice considered here, the phase diagram has been the subject of several studies for some specific theories \cite{Burnell11_2,Schulz13,Schulz14,Dusuel15,Schotte19}. 
For $J_\mathrm{l}=0$, the system is in a topological (string-net condensed \cite{Levin05}) phase ${\rm D} \mathcal C$ with a ground-state degeneracy that depends on the surface topology and excitations that are fluxons. By contrast, for $J_\mathrm{p}=0$, the system is in a trivial (non topological) phase with a unique ground state (all links in the trivial state $s=1$) and excitations that are link configurations with nontrivial strings satisfying the branching rules. These two phases are separated by a transition point that depends on the theory considered. In two dimensions, for Abelian theories ($\mathbb{Z}_N$ fusion rules), this model has been shown to be equivalent to the quantum Potts model in a transverse field defined on the dual (triangular) lattice \cite{Burnell11_2}, so that the transition is second-order for $N=2$ and first-order for $N \geqslant 3$. For non-Abelian theories, the situation is less clear. First studies based on series expansions and exact diagonalizations indicate a scenario compatible with second-order transitions (at least for Fibonacci \cite{Schulz13} and Ising theories \cite{Schulz14}), but the latest mean-field \cite{Dusuel15} and tensor-network approaches \cite{Schotte19} rather plead in favor of first-order transitions for all cases. \\

%
%
%%%%%%%%%%%%%%%%%%%
%%%%%%%%%%%%%%%%%%%
\noindent\emph{Wegner-Wilson loops.}---
%%%%%%%%%%%%%%%%%%%
%%%%%%%%%%%%%%%%%%%
%
%
The Hamiltonian (\ref{eq:ham}) may be seen as a generalization of lattice gauge theories. Indeed, when the input theory is associated to a group, $H$ describes a pure gauge theory (no matter) and the transition between the topological and the trivial phase driven by the fluxon dynamics  is a deconfinement/confinement transition of the charge excitations. As early proposed~\cite{Wegner71,Wilson74}, this transition is associated with a change of behavior of the Wegner-Wilson loops that exhibit a perimeter law in the deconfined (topological) phase and an area law in the confined (trivial) phase (see discussion below). The tension of these closed loops informs one about the interaction energy between the excitations existing at the extremities of the corresponding open strings. For instance, in the $\mathbb{Z}_2$ case, the closed string obtained by creating and annihilating a pair of electric charges and that measures the magnetic flux inside the resulting region, indicates that $J_{\rm l}$ is responsible for the charge confinement while fluxons condense (see, e.g., Ref.~\cite{Sachdev19} for more discussions). 

Hence, it is of crucial importance to determine the behavior of the loops. In their original paper,  Levin and Wen give the procedure to compute the matrix elements of the Wegner-Wilson loops in the link basis \cite{Levin05}. As explained in Ref.~\cite{,Burnell10}, these expressions are given in terms of $F-$symbols and $R-$symbols of the input theory. In the fluxon basis, the Wegner-Wilson loop $W^{(s,s')}_\C$ obtained by creating, moving, and annihilating a pair of excitations $(s,s')$ around a given region $\C$ of the lattice, is simply represented by two closed strings, $s$ and $s'$, above and below this region as depicted in  Fig.~\ref{fig:loops}. In this representation, these loops can be deformed at will, provided one forbids crossings with nontrivial strings and with links of the honeycomb lattice. 
%Indeed, consider three regions $\C_1, \C_2$, and $\C_3$ such that $\C_3=\C_1 \cup \C_2$. Then, consider an arbitrary $|\psi \rangle$ with no fluxons in the region $\C_2$, one has:
%\be
%W^{(s,s')}_{\C_1} \prod_{p \in \C_2} B_p |\psi \rangle =W^{(s,s')}_{\C_3} \prod_{p \in \C_2} B_p |\psi \rangle,
%\ee 
% so that $W^{(s,s')}_{\C_1} |\psi \rangle=W^{(s,s')}_{\C_3} |\psi \rangle$. 
 Thus, for $J_{\rm l}=0$, one simply has to evaluate a diagram with loops of type $s$ and $s'$ above and below the lattice in the ground state (no-fluxon state) of the Levin-Wen Hamiltonian. Using graphical rules \cite{Kitaev03,Bonderson_thesis}, this  directly leads to $\big \langle W^{(s,s')}_\C \big \rangle=d_s d_{s'}$. As explained above, the ground-state degeneracy in the topological phase depends on the surface topology. The results given here and below are valid for any ground states as long as the region $\C$ is contractible. \\

%
%
%%%%%%%%%%%%%%%%%%%
%%%%%%%%%%%%%%%%%%%
\noindent\emph{The weak-tension limit: $J_{\rm l} \ll J_{\rm p}$.}---
%%%%%%%%%%%%%%%%%%%
%%%%%%%%%%%%%%%%%%%
%
%
To compute $\big \langle W^{(s,s')}_\C \big \rangle$ in this limit, we use the same method as the one used for the toric code in a magnetic field in Ref.~\cite{Halasz12}. This approach, based on the perturbative continuous unitary transformations (PCUT) \cite{Wegner94,Glazek93,Glazek94,Stein97,Knetter00,Knetter03_1}, provides a clear picture of the various processes contributing to the perturbative corrections. Technically, this perturbative calculation amounts to evaluate diagrams corresponding to virtual excitations [see Fig.~\ref{fig:fundamental} (right), for example].
For simplicity, we consider here Wegner-Wilson loops defined on a hexagonal-shape closed region $\C$ (see Fig.~\ref{fig:loops}). In the perturbative limit where  \mbox{$\lambda=J_{\rm l}/J_{\rm p} \ll 1$} and for sufficiently large $\C$, one gets the general structure 
%
%
%%%%%%%%%%%%%%
 \be
\big \langle W^{(s,s')}_\C \big \rangle = d_s d_{s'}+  \sum_{n>0} \lambda^n w_n(L),
\label{eq:W_lf_general}
 \ee
 %%%%%%%%%%%%%%
 %
 %
 where $w_n$'s are polynomial of order $2\times  \lfloor n/2 \rfloor$, and $L$ is the number of links defining the contour of $\C$ (dimensionless perimeter).  The first terms of this expansion up to order $\lambda^4$ read
%
%
 %%%%%%%%%%%%%%
 \beqn
w_1(L)&=&0, \label{eq:5} \\
w_2(L)&=&-({d_s d_{s'}}-\delta_{s,s'})\frac{L}{4 D^2} ,\\
w_3(L)&=&-({d_s d_{s'}}-\delta_{s,s'})\frac{L(D^2+2)}{4 D^4},\\
w_4(L)&=&  ({d_s d_{s'}}-\delta_{s,s'})\bigg\{\frac{d_s d_{s'} (94+7 D^2)+9 \, \delta_{s,s'}}{24 \,d_s d_{s'} D^6} \nonumber\\ 
 &&+ \frac{L[d_s d_{s'}(86-331D^2-18 D^4)+9 \, \delta_{s,s'}]}{96 \,d_s d_{s'} D^6} \nonumber\\
 &&+({d_s d_{s'}}-\delta_{s,s'}) \frac{L^2}{32 D^4}\bigg\}.  \label{eq:8} 
\eeqn
%%%%%%%%%%%%%%
 %
 %
 Calculation details will be given elsewhere \cite{Vidal21}. These expressions suggest that, in the topological phase, Wegner-Wilson loops obey a perimeter law, i.e., $\big \langle W^{(s,s')}_\C \big \rangle \propto {\rm e}^{-\#L}$, expected for deconfined phases~\cite{Kogut79} (see Ref.~\cite{Halasz12} for a similar exponentiation). \\

%
%
%%%%%%%%%%%%%
%%%%%%%%%%%%%
%
%

%
%
%%%%%%%%%%%%%%%%%%%
%%%%%%%%%%%%%%%%%%%
\noindent\emph{The strong-tension limit: $J_{\rm l} \gg J_{\rm p}$.}---
%%%%%%%%%%%%%%%%%%%
%%%%%%%%%%%%%%%%%%%
%
%
In this other limiting case, the behavior of the Wegner-Wilson loops is completely different and it is more appropriate to work in the original link basis. For $J_{\rm p}=0$, keeping in mind that the ground state is the product state $\otimes_l |1\rangle_l$,  \mbox{where $|1\rangle_l$} denotes the state $s=1$ in the link $l$, one straightforwardly gets $\big \langle W^{(s,s')}_\C \big \rangle=\delta_{s,s'}$. However, contrary to the topological phase, the first nontrivial contribution occurs at order $(1/\lambda)^A$, where $A$ is the number of plaquettes inside the region $\C$ (dimensionless area). More precisely, one has:
%
%
%%%%%%%%%%%%
\beqn
\big \langle W^{(s,s')}_\C \big \rangle&=&\delta_{s,s'} + \gamma_\C ({d_s d_{s'}}-\delta_{s,s'})\left(\frac{1}{\lambda D^2}\right)^A \nonumber \\
&&+\,O(1/\lambda^{A+1}),
\label{eq:W_pertur_st}
\eeqn
%%%%%%%%%%%%
%
%
where $\gamma_\C$ is a purely combinatorial factor that depends on the region $\C$ but not on the theory $\mathcal{C}$. For instance, if $\C$ consists in two adjacent plaquettes, one has $\gamma_\C=11/90$. This behavior can be interpreted as an area law for the quantity $\big \langle W^{(s,s')}_\C \big \rangle-\delta_{s,s'} \propto {\rm e}^{-\#A}$.\\

%
%
%%%%%%%%%%%%%%%%%%%
%%%%%%%%%%%%%%%%%%%
\noindent\emph{Mean-field approach.}---
%%%%%%%%%%%%%%%%%%%
%%%%%%%%%%%%%%%%%%%
%
%
It is interesting to compare the results obtained perturbatively with the ones computed from the mean-field ansatz introduced in Ref.~\cite{Dusuel15}
%
%
%%%%%%%%%%%%%%%%
\begin{equation}
\ket{\alpha}=\mN \prod_p (\id+\alpha Z_p) \otimes_l |1\rangle_l,
\label{eq:state}
\end{equation}
%%%%%%%%%%%%%%%%
%
%
where $\mN$ is the normalization constant, $0\leqslant\alpha\leqslant 1$ is a variational parameter, and $Z_p=2\Pp-\id$. This variational state, which interpolates between one exact ground state for $J_\mathrm{p}=0 \, (\alpha=0)$, and the exact ground state for $J_\mathrm{l}=0\,  (\alpha=1)$, leads to 
%
%
%%%%%%%%%%%%
\be
\big \langle W^{(s,s')}_\C  \big \rangle_\alpha =\delta_{s,s'}+ ({d_s d_{s'}}-\delta_{s,s'}) \left(\frac{D^2\langle \Pp \rangle_\alpha-1}{D^2-1}\right)^A,
\label{eq:mf}
\ee
%%%%%%%%%%%%
%
%
where $\langle\mathcal{O}\rangle_\alpha=\bra{\alpha}\mathcal{O}\ket{\alpha}$. This mean-field approach relies on a description in terms of decoupled plaquettes which is encoded in the following factorization property~\cite{Dusuel15}
%
%%%%%%%%%%%%%%%%
\begin{equation}
\Big\langle \prod_{p \in \C}  \Pp \Big\rangle_\alpha =\prod_{p \in \C} \langle \Pp \rangle_\alpha=\left[\frac{(1+\alpha)^2}{D^2(1-\alpha)^2+4\alpha}\right]^A.
\label{eq:factorization}
\end{equation}
%%%%%%%%%%%%%%%%
%
%

In the topological phase, one has $\alpha=1$ and  \mbox{$\langle \Pp \rangle_\alpha=1$}~\cite{Dusuel15}. Thus, this ansatz yields a trivial perimeter law which corresponds to the leading-order contribution in $\lambda^0$ given in Eq.~(\ref{eq:W_lf_general}), i.e.,  \mbox{$\big \langle W^{(s,s')}_\C  \big \rangle_\alpha=d_s d_{s'}$}. 
By contrast, in the strong-tension limit (trivial phase), one finds:
%
%
%%%%%%%%%%%%
\beqn
\alpha&=&\frac{1}{12 \lambda}+O(1/\lambda^2),\\
\langle \Pp \rangle_\alpha&=&\frac{1}{D^2}+\frac{1}{\lambda}\frac{D^2-1}{3 D^2}+O(1/\lambda^2),\\
\big \langle W^{(s,s')}_\C  \big \rangle_\alpha&=&\delta_{s,s'}+\frac{1}{3^A} (d_s d_{s'}-\delta_{s,s'}) \left(\frac{1}{\lambda D^2}\right)^A \nonumber \\
&&+O(1/\lambda^{A+1}).
\label{eq:W_mf}
\eeqn
%%%%%%%%%%%%
%
%
Hence, it is remarkable to observe that the mean-field result (\ref{eq:W_mf}) reproduces the perturbative result (\ref{eq:W_pertur_st}) with a factor $\gamma_\C=1/3^A$ that only depends on the area  (but not on the shape) of $\C$ within this approximation. \\

%
%
%%%%%%%%%%%%%%%%%%%
%%%%%%%%%%%%%%%%%%%
\noindent\emph{Discussion.}---
%%%%%%%%%%%%%%%%%%%
%%%%%%%%%%%%%%%%%%%
%
%
Let us now discuss the results that can be inferred from the perturbative calculations. Three cases must be distinguished according to the nature of the strings defining the Wegner-Wilson loops $W^{(s,s')}_\C$:

\begin{enumerate}

\item If $s'=s$ and $d_s=1$, one can use the following identity (valid for any Abelian strings $s$ and $s'$):
%
%
%%%%%%%%%%%%%%%%
\begin{equation}
W^{(s,s')}_\C =\prod_{p\in \C} B_p^{s} \big(B_p^{s'}\big)^\dagger=\prod_{p\in \C} B_p^{s} B_p^{\overline{s'}}= \prod_{p\in \C} B_p^{s \times \overline{s'}},
\label{eq:product}
\end{equation}
%%%%%%%%%%%%%%%%
%
%
to show that $W^{(s,s)}_\C=\id$. Here,  $\overline{s}$ denotes the dual (or conjugate) string of $s$, i.e., $1\in s\times \overline{s}$ \cite{Levin05}. This is in agreement with the perturbative results given in \mbox{Eqs.~(\ref{eq:5})-(\ref{eq:W_pertur_st})} as well as the mean-field approach [see Eq.~(\ref{eq:mf})]. We conclude that Abelian fluxons are always completely deconfined. In other words,  $\big \langle W^{(s,s)}_\C \big \rangle=1$, for all $\lambda$. 

\item If $s=s'$ and $d_s>1$ (non-Abelian strings), Eq.~(\ref{eq:product}) is not valid. In this case, one gets a perimeter law in the topological phase and a modified  area law in the trivial phase. Indeed, Eq.~(\ref{eq:W_pertur_st}) indicates that $\displaystyle{\lim_{A \rightarrow \infty}\big \langle W^{(s,s)}_\C \big \rangle=1}$, which is remi\-niscent of a deconfinement of excitations $(s,s)$ in the trivial phase.

\item If $s \neq s'$, $\big \langle W^{(s,s')}_\C \big \rangle/(d_s d_{s'})$  depends only on $D$ and $\C$ (at least at the order considered here) and obeys a perimeter law in the topological phase and an area law in the trivial phase.

\end{enumerate}

%
%
%%%%%%%%%%%%%
%%%%%%%%%%%%%
%
%

%
%
%%%%%%%%%%%%%%%%%%%
%%%%%%%%%%%%%%%%%%%
\noindent\emph{Perspectives.}---
%%%%%%%%%%%%%%%%%%%
%%%%%%%%%%%%%%%%%%%
%
%
To go beyond the present work, several extensions should be considered. Concerning input theories, one may study the case of UMTCs with nontrivial Frobenius-Schur indicators such as semion or $SU(2)_2$ theories \cite{Rowell09}, and/or with nontrivial multiplicities. Input theories that are not UMTCs are also of interest. In that respect, the simplest example is the $\mathbb{Z}_2$ gauge theory for which there are four Wegner-Wilson loops labeled $W_{i=1,..,4}$ in Ref.~\cite{Levin05}. It turns out that, in the charge-free sector, $\langle W_1\rangle =\langle W_3 \rangle=1$, and the expression of $\langle W_2 \rangle=\langle W_4 \rangle$ can be obtained from \mbox{Eqs.~(\ref{eq:5})-(\ref{eq:W_pertur_st})} by setting $d_s=d_{s'}=1$, $D=\sqrt{2}$, and $\delta_{s,s'}=0$. More generally, discrete gauge theories associated to non-Abelian gauge groups (see Ref.~\cite{Levin05} for a concrete example based on the $S_3$ group) definitely deserve special attention.\\

%%%%%%%%%%%%%%%%%%%
%%%%%%%%%%%%%%%%%%%

\acknowledgments
We thank S. Dusuel and M. Tissier for fruitful discussions. We are also grateful to  M. M\"uhlhauser and K. P. Schmidt for providing the PCUT coefficients.\\


%\bibliography{biblio_WWLW.bib}

\begin{thebibliography}{47}
\bibitem [{Not()}]{Note}%
  \BibitemOpen
  \href@noop {} {}\bibinfo {note} {{Here, we consider only multiplicity-free
  UMTCs whose labels have trivial Frobenius-Schur indicators.}}\BibitemShut
  {Stop}%
\end{thebibliography}

\end{document}